# Virtual Network Embedding Algorithms Based on Best-Fit Subgraph Detection

Ashraf A. Shahin[1,2]

[1] College of Computer and Information Sciences, Al Imam Mohammad Ibn Saud Islamic University (IMSIU) Riyadh, Kingdom of Saudi Arabia

[2] Department of Computer and Information Sciences, Institute of Statistical Studies & Research, Cairo University, Egypt

Correspondence: Ashraf A. Shahin, College of Computer and Information Sciences, Al Imam Mohammad Ibn Saud Islamic University (IMSIU) Riyadh, Kingdom of Saudi Arabia. E-mail: ashraf_shahen@ccis.imamu.edu.sa

## Abstract

One of the main objectives of cloud computing providers is increasing the revenue of their cloud datacenters by accommodating virtual network requests as many as possible. However, arrival and departure of virtual network requests fragment physical network's resources and reduce the possibility of accepting more virtual network requests. To increase the number of virtual network requests accommodated by fragmented physical networks, we propose two virtual network embedding algorithms, which coarsen virtual networks using *Heavy Edge Matching* (HEM) technique and embed coarsened virtual networks on best-fit sub-substrate networks. The performance of the proposed algorithms are evaluated and compared with existing algorithms using extensive simulations, which show that the proposed algorithms increase the acceptance ratio and the revenue.

## 1. Introduction

Network virtualization is one of the most important features of the cloud computing, which allows creation of multiple virtual networks (VNs) on a single substrate network (SN) (Fajjari et al, 2013). In cloud computing, virtual network requests (VNRs) arrive dynamically over time with different topologies, resource requirements, and lifetime. Cloud providers accommodate VNRs by mapping each virtual node to a substrate node and each virtual link to a substrate path, such that a set of previously defined mapping constraints (e.g. topology constraints, link bandwidth, CPU capacity) is satisfied (Lischka & Karl, 2009). At the end of the VNR's lifetime, allocated SN's resources are released.

However, mapping virtual networks' resources to substrate network's resources is known to be nondeterministic polynomial-time hard (*NP-hard*) (Chowdhury et al, 2009; Chowdhury et al, 2012). This problem is usually referred to as the *Virtual Network Embedding* (VNE) problem (Till Beck et al, 2013). In the last few years, many studies have proposed several algorithms for efficient and effective VNE (Lischka & Karl, 2009; Chowdhury et al, 2009; Chowdhury et al, 2012; Cheng et al, 2011; Zhang et al, 2013; Cheng et al, 2012). Effective VNE increases the utilization of the SN's resources and increases the revenues of the cloud computing datacenters. However, the process of allocating and releasing SN's resources fragments SN's resources (Fischer et al, 2013). SN's resources are considered fragmented if there are two or more sub-substrate networks connected by substrate paths with available bandwidths less than the minimum required VNR's bandwidth or substrate paths with lengths greater than the maximum allowed hops.

Fragmentation of the SN's resources reduces the acceptance ratio and reduces the long-term revenue. Substrate resources fragmentation can be improved by re-embedding one or more VNRs (Fischer et al, 2013). However, re-embedding running VNRs may reduce the quality of service provided to the customers and may violate the service level agreement (SLA).

Most of current researches ignore the SN's resources fragmentation problem (Chowdhury et al, 2009; Cheng et al, 2011; Zhang et al, 2013), while others propose reactive algorithms (Zhu & Ammar, 2006; Fajjari et al, 2011).

Reactive algorithms just act when a VNR is rejected, and try to re-embed one or more VNRs to increase the acceptance ratio. Some of the current researches consider load balance of the SN's resources to avoid SN's resources fragmentation (Cheng et al, 2012; Sun et al, 2013). However, considering load balance of the SN's resources during embedding VNRs with small resources may cause rejecting VNRs with large resources in the future.

In this paper, we propose two VNE algorithms to improve the acceptability of fragmented substrate networks. The proposed algorithms increase the possibility of accepting future VNRs by embedding VNRs on best-fit sub-substrate networks. The proposed algorithms are one stage (embed virtual nodes and virtual links at the same stage to allow coordination between them), online (deal with VNRs that arrive over time and do not require VNRs to be previously known), and backtracking algorithms.

Both algorithms exploit the virtualization feature of the cloud computing to embed more than one virtual node from the same VN on the same substrate node, which minimizes the cost of embedding VNRs by eliminating the cost of embedding virtual links between those virtual nodes. Additionally, the second algorithm coarsens VNs using *Heavy Edge Matching* (HEM) technique to minimize its size and to minimize the total required bandwidth before embedding them on best-fit sub-substrate networks.

The performance of the proposed algorithms have been evaluated using extensive simulations, which show that the proposed algorithms increase the long-term average revenue and the acceptance ratio compared to some of the existing embedding algorithms.

The rest of this paper is organized as follows. In Section 2, we discuss the related work on VNE problem. Section 3 presents the VN embedding model and problem formulation. Section 4 describes the proposed algorithms. Section 5 evaluates the proposed VN embedding algorithms using extensive simulations. Finally, in Section 6 we conclude this paper.

## 2. Related Work

Many researches have been done for efficient VN embedding. Zhu and Ammar (Zhu & Ammar, 2006) proposed two VN embedding algorithms. The first algorithm is static VNE algorithm, where allocated substrate resources are fixed throughout the VN lifetime. Heuristics and adaptive optimization strategies are used to improve the performance of the proposed algorithm. The second algorithm reconfigures the embedded VNs to increase the utilization of the underlying substrate resources. However, the proposed algorithms deal only with offline embedding problem (all VNRs are previously known). On the other hand, in cloud computing data centers, VNE problem is online problem, where new VNRs arrive over time.

Lischka and Karl (Lischka & Karl, 2009) presented online backtracking VNE algorithm. The proposed algorithm is one stage VNE algorithm (maps nodes and links during the same stage) and maps VN to a sub-physical network that is similar to the topology of the VN and achieves previously defined constraints (e.g. CPU capacity, link bandwidth). Nodes are mapped to substrate nodes sequentially after sorting it in descending order based on its required CPU. The proposed algorithm tries to map virtual links to substrate paths with minimal hops by incrementally increasing the maximum hop limit. However, the computational complexity is high due to multiple operations of the proposed algorithm. Di et al. (Di et al, 2010) improved performance and complexity of the proposed algorithm in (Lischka & Karl, 2009) by considering link mapping cost during the process of sorting nodes and choosing the maximal hop limit. Nogueira et al. (Nogueira et al, 2011) proposed heuristic-based VN embedding algorithm to deal with the heterogeneity of VNs and SN, in both links and nodes. The proposed algorithm is one stage VNE algorithm.

Cheng et al. (Cheng et al, 2011) proposed two online virtual network-embedding algorithms called RWMaxMatch and RW-BFS. Both of them rank nodes using topology-aware node ranking technique to reflect the topological structure of the VNs and the SN. RWMaxMatch algorithm is two stage embedding algorithm, which performs node mappings and link mappings at two different stages without coordination between them. However, mapping nodes without considering its relation to the link mapping might result in neighboring virtual nodes being widely separated in the SN, which leads to high consumption of the underlying SN's resources and reduces the acceptance ratio of the proposed algorithm. To avoid this problem, they proposed RW-BFS algorithm. RW-BFS algorithm is a backtracking one stage embedding algorithm, which embeds nodes and links at the same stage.

To improve the coordination between nodes mapping stage and links mapping stage, Chowdhury et al. (Chowdhury et al, 2009; Chowdhury et al, 2012) formulated the VNE problem as a mixed integer program (MIP), which is NP-hard. To obtain polynomial-time solvable algorithms, they relaxed the integer program to linear program, and proposed two VNE algorithms: D-ViNE (deterministic VNE algorithm) and R-ViNE (randomized VNE algorithm).

Zhang et al. (Zhang et al, 2013) proposed two VN embedding models: an integer linear programming model and a mixed integer programming model. To solve these models, the authors proposed VN embedding algorithm based on particle swarm optimization. The time complexity of the link mapping stage is reduced by using shortest path algorithm and greedy k-shortest paths algorithm.

Some of existing works proposed VN embedding algorithms to embed VNRs in distributed cloud computing environments (Samuel et al, 2013; Houidi et al, 2008a; Xin et al, 2011; Lv et al, 2010). Houidi et al. (Houidi et al, 2011) proposed exact and heuristics VN embedding algorithms, which split virtual network requests using max-flow min-cut algorithms and linear programming techniques. Leivadeas et al. (Leivadeas et al, 2013) proposed VN embedding algorithm based on linear programming. The proposed algorithm partitions VNRs using partitioning approach based on Iterated Local Search. Houidi et al. (Houidi et al, 2008b) proposed distributed VN embedding algorithm, which is performed by agent-based substrate nodes. The authors proposed VN embedding protocol to allow communication between the agent-based substrate nodes. However, the proposed algorithm assumes that all VNRs are previously known.

### 3. Virtual Network Embedding Model and Problem Formulation

***Substrate network (SN):*** We model the substrate network as a weighted undirected graph $G_s = (N_s, L_s)$, where $N_s$ is the set of substrate nodes and $L_s$ is the set of substrate links. Each substrate node $n_s \in N_s$ is weighted by the CPU capacity, and each substrate link $l_s \in L_s$ is weighted by the bandwidth capacity. Figure 1 (b) shows a simple SN example, where the available CPU resources are represented by numbers in rectangles and the available bandwidths are represented by numbers over the links.

***Virtual network (VN):*** virtual network $VN_i$ is modeled as a weighted undirected graph $G_{v_i} = (N_{v_i}, L_{v_i})$, where $N_{v_i}$ is the set of virtual nodes and $L_{v_i}$ is the set of virtual links. Virtual nodes and virtual links are weighted by the required CPU and bandwidth, respectively. Figure 1 (a) shows an example of VN with required CPU and bandwidth.

***Virtual network requests (VNR):*** the $i^{th}$ VN request $vnr_i$ in the set of all VN requests $VNR$ is modeled as $(G_{v_i}, t_{a_i}, t_{l_i})$, where $G_{v_i}$ is the required VN to be embedded, $t_{a_i}$ is the arrival time, and $t_{l_i}$ is the lifetime. When $vnr_i$ arrives, substrate nodes' CPU and substrate links' bandwidth are allocated to achieve the $vnr_i$. If the substrate network does not have enough resources to achieve $vnr_i$, $vnr_i$ is rejected. At the end of $vnr_i$ lifetime, all allocated resources to $vnr_i$ are released.

***Virtual Network Embedding (VNE):*** embedding $VN_i$ on SN is defined as a map $M: G_{v_i} \rightarrow (N_s', P_s')$, where $N_s' \subseteq N_s$, and $P_s' \subseteq P_s$ , where $P_s$ is the set of all loop free substrate paths in $G_s$. Embedding $VN_i$ can be decomposed into node and link mapping as follows:

$$\text{Node mapping: } M_N : N_{v_i} \rightarrow N_s'$$

$$\text{Link mapping: } M_L : L_{v_i} \rightarrow P_s'$$

For example, mapping of the VN in figure 1 (a) on SN in figure 1(b) can be decomposed into:

$$\text{Node mapping: } \{a \rightarrow B, b \rightarrow A, c \rightarrow E\}$$

$$\text{Link mapping: } \{(a,b) \rightarrow \{(B,A)\}, (b,c) \rightarrow \{(A,D),(D,E)\}, (c,a) \rightarrow \{(E,B)\}\}$$

***Virtual Network Embedding Revenue:*** as in (Cheng et al, 2011; Zhang et al, 2013; Fischer et al, 2013), the revenue of embedding $vnr_i$ at time $t$ is defined as the sum of all required substrate CPU and substrate bandwidth by $vnr_i$ at time $t$.

$$R(vnr_i, t) = Life(vnr_i, t). \left( \sum_{n_{v_i} \in N_{v_i}} CPU(n_{v_i}) + \sum_{l_{v_i} \in L_{v_i}} BW(l_{v_i}) \right)$$

Where $CPU(n_{v_i})$ is the required CPU for the virtual node $n_{v_i}$, $BW(l_{v_i})$ is the required bandwidth for the virtual link $l_{v_i}$, and $Life(vnr_i, t) = 1$ if $vnr_i$ is in its lifetime and substrate resources are allocated to it, otherwise $Life(G_{v_i}, t) = 0$.

***Virtual Network Embedding Cost:*** as in (Cheng et al, 2011; Zhang et al, 2013; Fischer et al, 2013), the cost of embedding $vnr_i$ at time $t$ is defined as the sum of all allocated substrate CPU and substrate bandwidth to $vnr_i$ at time $t$.

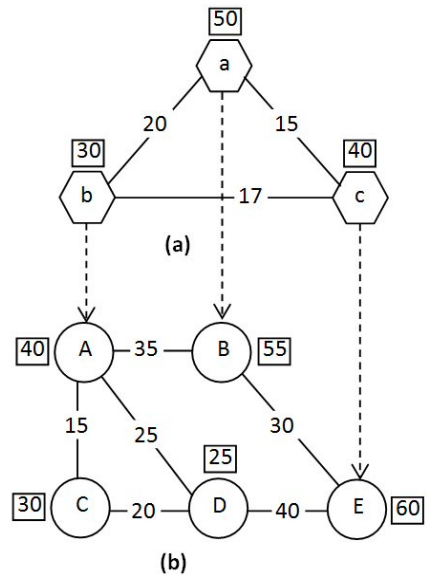

Figure 1. Example of VNE

$$Cost(vnr_i, t) = Life(vnr_i, t).\left(\sum_{n_{v_i} \in N_{v_i}} CPU(n_{v_i}) + \sum_{l_{v_i} \in L_{v_i}} BW(l_{v_i}).Length(M_{L_{v_i}}(l_{v_i}))\right)$$

Where $Length(M_{L_{v_i}}(l_{v_i}))$ is the length of the substrate path that the virtual link $l_{v_i}$ is mapped to.

***Objectives***: the main objectives are to increase the revenue and decrease the cost of embedding virtual networks in the long run. To evaluate the achievement of these objectives, we use the following metrics:

- *The long-term average revenue*, which is defined by

$$\lim_{T \to \infty} \left(\frac{\sum_{t=0}^{T} \sum_{i=1}^{I} R(vnr_i, t)}{T}\right) \qquad (1)$$

Where $I = \| VNR \|$, and $T$ is the total time.

- *The VNR acceptance ratio*, which is defined by

$$\frac{\|VNR_s\|}{\|VNR\|} \qquad (2)$$

Where $VNR_s$ is the set of all accepted virtual network requests.

- *The long term R/Cost ratio*, which is defined by

$$\lim_{T \to \infty} \left(\frac{\sum_{t=0}^{T} \sum_{i=1}^{I} R(vnr_i, t)}{\sum_{t=0}^{T} \sum_{i=1}^{I} Cost(vnr_i, t)}\right) \qquad (3)$$

## 4. The Proposed Algorithms

We proposed two VNE algorithms to increase the acceptance ratio for fragmented SNs and minimize the remapping process. In the next sub-sections, we describe the motivation behind the proposed algorithms and describe the details of the proposed algorithms.

### 4.1 Motivation

For each VNR, we may have different VNE solutions. For example, table 1 shows some VNE solutions to embed VN in figure 2(a) on the SN in figure 2(c). Although, solution 2 consumes less SN's resources among others, it reduces the possibility to accept more future VN requests, such as VN in figure 2(b). Mapping VN in figure 2(b) on the SN in figure 2(c) requires remapping the VN in figure 2(a) to solution1. Then, the VN in figure 2(b) can be mapped as following:

Node mapping: $\{d \to E, e \to J, f \to D, g \to H\}$

Link mapping: $\{(d, e) \to \{(E, J)\}, (e, g) \to \{(J, I), (I, H)\}, (g, f) \to \{(H, D)\}, (f, d) \to \{(D, E)\}\}$

From this example, we conclude that increasing the possibility to accept more future VN requests does not only depend on the mapping cost. This conclusion motivates us to propose VNE algorithms for selecting VNE solutions that increase the possibility to accept as many VNRs as possible. Even if the selected solutions maybe increase the embedding cost, the revenue is increased by accommodating more VNRs.

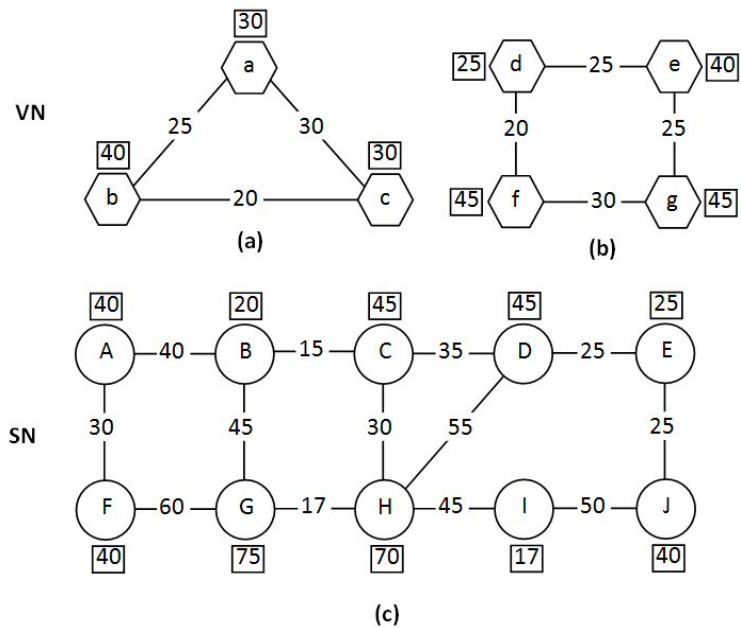

Figure 2. Example of VNs and SN

Table 1. Some VNE solutions to embed VN in figure 2(a) on the SN in figure 2(c)

| | Node mapping | Link mapping |
|---|---|---|
| Solution1 | $\{a \rightarrow A, b \rightarrow F, c \rightarrow G\}$ | $\{(a,b) \rightarrow \{(A,F)\}, (b,c) \rightarrow \{(F,G)\}, (c,a) \rightarrow \{(G,B),(B,A)\}\}$ |
| Solution2 | $\{a \rightarrow C, b \rightarrow H, c \rightarrow D\}$ | $\{(a,b) \rightarrow \{(C,H)\}, (b,c) \rightarrow \{(H,D)\}, (c,a) \rightarrow \{(D,C)\}\}$ |
| Solution3 | $\{a \rightarrow D, b \rightarrow J, c \rightarrow H\}$ | $\{(a,b) \rightarrow \{(D,E),(E,J)\}, (b,c) \rightarrow \{(J,I),(I,H)\}, (c,a) \rightarrow \{(H,D)\}\}$ |

*4.2 BFSN VNE Algorithm*

The inputs to BFSN algorithm are: (1) $G_v$: the virtual network to be embed, (2) $G_s$: the substrate network to embed on, (3) Max_hops: the maximum allowed substrate path length, and (4) Max_backtrack: the upper bound of nodes re-mapping operation. The outputs of BFSN algorithm is a map $M(G_v)$.

Figure 3 shows the steps of BFSN algorithm. The algorithm constructs breadth-first searching tree of $G_v$. The root node of the constructed tree is the virtual node with the largest resources (sum of CPU and BW). Nodes in each level in the created breadth-first searching tree are sorted in descending order based on their resources.

The set of candidate sub-substrate networks $G_{sub} = \{G_{sub_i} \mid G_{sub_i} = (N'_{s_i}, P'_{s_i}), N'_{s_i} \subseteq N'_s, P'_{s_i} \subseteq P'_s\}$ is built, where $N'_s$ is a set of all substrate nodes in $G_s$ with available CPU $\geq$ minimum required CPU (the minimum CPU in $G_v$), and $P'_s$ is the set of all loop free substrate paths in $G_s$ with available BW $\geq$ minimum required BW and with length less than or equal a pre-specified maximum path length.

Candidate sub-substrate networks are specified by visiting substrate nodes in $N'_s$ sequentially and creating a breadth-first searching tree from a substrate node if it has not been included in any sub-substrate network yet. Substrate node is added to the sub-substrate network if it has not been included in any sub-substrate network, and has been connected with previously added substrate node through a substrate path in $P'_s$. $G_{sub}$ is reduced by removing sub-substrate networks that do not have enough resources to embed $G_{v_i}$. All remaining sub-substrate networks in $G_{sub}$ are sorted in ascending order based on their total available resources.

The BFSN algorithm embeds VN on a sub-SN using *Embed()* function, which is described in figure 6. In the *Embed()* function, candidate substrate node list for each virtual node is built by collecting all substrate nodes that have available CPU capacity at least as large as the virtual node CPU and have a loop free substrate path to each substrate node contains one of the previously mapped neighbors. Each substrate path should satisfy the constraint of the maximum substrate path length, and have available bandwidth greater than or equal the bandwidth of the virtual link between the virtual node and its previously mapped neighbor.

Candidate substrate nodes for each virtual node are collected by creating a breadth-first search tree from each substrate node contains one of the previously mapped neighbors, and finding the common substrate nodes between the created trees. In the constructed trees, substrate nodes should satisfy the CPU constraints for virtual node, and substrate paths should satisfy the connectivity constraints to connect the virtual node with its neighbors. By this way, all candidate substrate nodes in the list satisfy all constraints (CPU and connectivity constraints).

Substrate nodes in the candidate substrate node list are sorted in ascending order according to the total cost of embedding virtual links from the virtual node to all previously embedded neighbors. If the virtual node is a root node, the candidate substrate node list is a set of all substrate nodes that have enough resources to embed the virtual node. The candidate substrate nodes for the root are sorted in descending order according to the total available resources.

Virtual node is sequentially mapped to substrate nodes in its candidate substrate node list. If there is no appropriate substrate node in its candidate substrate node list, we backtrack to the previously mapped node, re-map it to the next candidate substrate node, and continue to the next node. Mappings of the virtual node and its virtual links are added to $M(G_v)$ by using the function *Add()* in line 3. In line 6, *Delete()* function is used to perform the backtracking process.

---

**Algorithm 1 The Details of the BFSN VNE algorithm**

---

1: Build breadth-first searching tree of $G_v$ from virtual node with largest resources.

2: Sort all nodes in each level in the created breadth-first tree in descending order according to their required resources.

3: Build candidate sub-substrate networks $G_{sub}$

4: for each $G_{sub_i} \in G_{sub}$ do

5: backtrack count=0

6:   if Embed($G_{v_{root}}$ , $G_{sub_i}$ , $M(G_v)$)  then

7:     return true

8: end for

9: return false

---

---

**Algorithm 2 The Details of Embed($n_{v_i}$ , $G_{sub_i}$, $M(G_v)$)  Function**

---

1: Build candidate substrate node list $C_i$ for $n_{v_i}$

2: for each $n_s$ in $C_i$

3:   Add $\left( (n_{v_i} , n_s) , M(G_v) \right)$

4:   if Embed($n_{v_{i+1}}$ , $G_{sub_i}$ , $M(G_v)$)  then return true

5:   else

6:     Delete$\left( (n_{v_i} , n_s) , M(G_v) \right)$

7:     if backtrack count > Max_backtrack then return false

8: end for

9: backtrack count ++

10: return false

---

For example, to map VN in figure 2 (a) on the SN in figure 2 (c), we construct a breadth-first searching tree from the virtual node *b*, which is the virtual node with largest resources. Virtual nodes in each level are sorted in descending order based on their resources as in figure 3. Figure 4 shows candidate sub-substrate networks that are specified, and sorted in ascending order based on its total available resources. Numbers in triangles represent numbers of hops for links. Figure 5 shows the embedding steps of the VN in figure 2 (a) on the first candidate sub-SN, which is the candidate sub-SN in figure 4 (a).

The candidate substrate nodes list for the root node *b* is built as $\{G, F, A\}$. The root node *b* is mapped to the substrate node *G* as shown in figure 5(b). The next virtual node in the breadth first search tree in figure 3 is the virtual node *a*. The set of candidate substrate nodes for the virtual node *a* is specified as $\{G, F, A\}$, which is sorted in ascending order based on the cost of mapping the virtual node *a* to each of them. *G* substrate node is included because the proposed algorithm allows substrate nodes to include more than one virtual node from the same VN. Figure 5(c) shows the SN after mapping the virtual node *a* to the substrate node *G*. Finally, the virtual node c is mapped to the substrate node F, which is the only substrate node that can satisfy the connectivity constraints to the substrate node G (the total required bandwidth is 50).

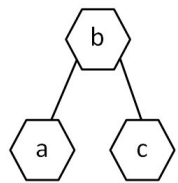

Figure 3. Sorted breadth-first searching tree for the VN in figure 2 (a)

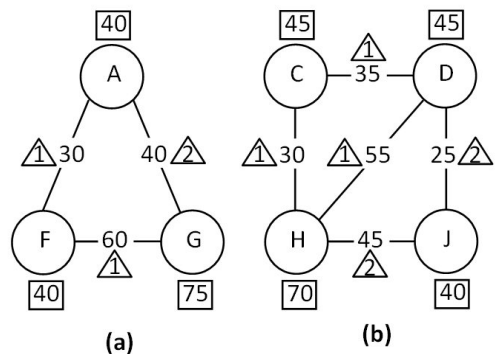

Figure 4. Candidate sub-substrate networks for the VN in figure 2 (a)

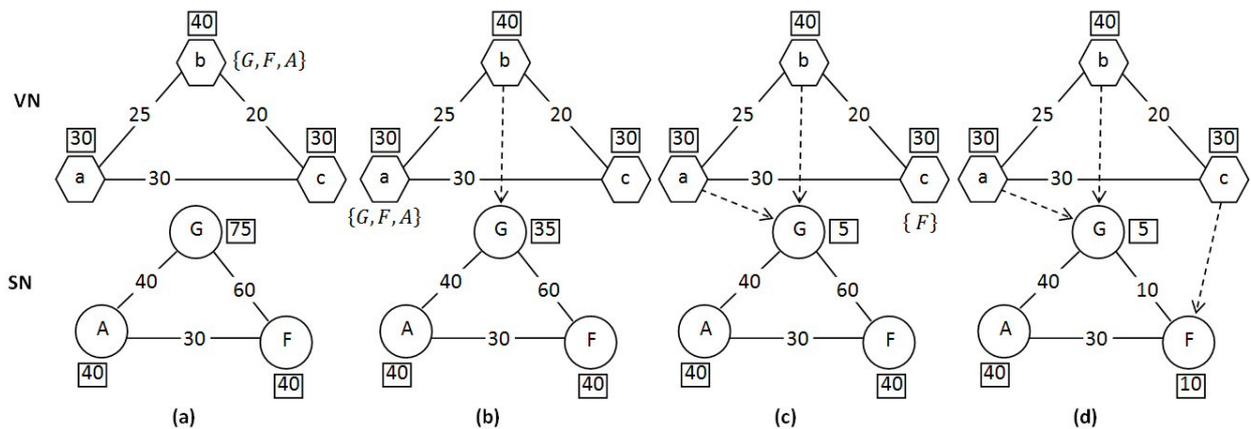

Figure 5. Embedding VN in figure 2 (a) on the candidate sub-substrate network in figure 4 (a)

*4.2 BFSN-HEM VNE Algorithm*

Allowing substrate nodes to contain more than one virtual node from the same VN coarsens the VN. For example, the VN in figure 2 (a) is coarsened as shown in figure 6. Coarsening VNs reduces the embedding cost by eliminating the cost of embedding virtual links between virtual nodes on the same substrate node. However, VNs coarsening performed by BFSN algorithm is not the optimal coarsening solution. For this reason, we proposed BFSN-HEM algorithm, which coarsens VNs using *Heavy Edge Matching* (HEM) technique to minimize its sizes and to minimize the edge-cut before embedding them. The Details of the *Coarsening()* function are shown in algorithm 3.

*Coarsening()* function sorts all links in descending order based on their bandwidth. The incident nodes of each link is merged if the total CPU less than or equal the maximum allowed CPU. After each coarsening step, VN is updated to a new coarsened virtual network with merged nodes and links. In the BFSN-HEM algorithm, coarsened virtual nodes are mapped as in BFSN algorithm. Coarsened virtual links are mapped by mapping its virtual links independently.

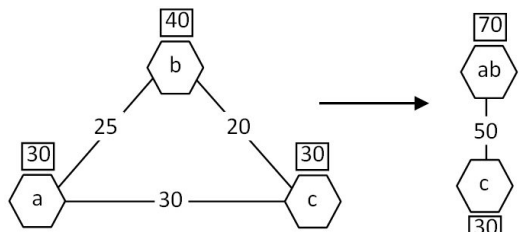

Figure 6. Example of VN coarsening

---

Algorithm 3 The Details of the Coarsening($G_v, CPU_{Max}$) function

---

1: Sort all links in $L_v$ in descending order based on their bandwidth.

2: for each $l_v \in L_v$ do

3:     if the summation of its incident nodes $\leq CPU_{Max}$ then

4:         Update $G_v$ by merging the incident nodes of $l_v$

5:         Coarsening($G_v, CPU_{Max}$)

6:         return $G_v$

7:     end if

8: end for

9: return $G_v$

---

## 5. Performance Evaluation

The proposed VN embedding algorithms are evaluated by comparing its performance with some of existing algorithms. First, we implemented four algorithms: BFSN, BFSN-HEM, RW-MaxMatch (Cheng et al, 2012), and RW-BFS (Cheng et al, 2011). Second, we generated SN topology and 3000 VN topologies to be used as inputs to the implemented algorithms. Finally, we compared the results from the implemented algorithms. In the following sub-sections, we describe the evaluation environment settings and discuss the results of the simulations.

*5.1 Evaluation Environment Settings*

In our evaluation, the substrate network topology is configured to have 200 nodes with 1000 links. Substrate network is generated using Waxman generator. Bandwidths of the substrate links are real numbers uniformly distributed between 50 and 100 with average 75. We have selected two server configurations: HP ProLiant ML110 G4 (Intel Xeon 3040, 2 cores X 1860 MHz, 4 GB), and HP ProLiant ML110 G5 (Intel Xeon 3075, 2 cores X 2660 MHz, 4 GB). Each substrate node is randomly assigned one of these server configurations.

Virtual network topologies are generated using Waxman generator with average connectivity 50%. Number of virtual nodes in each VN is variant from 2 to 20. Each virtual node is randomly assigned one of the following CPU: 2500 MIPS, 2000 MIPS, 1000 MIPS, and 500 MIPS, which are correspond to the CPU of Amazon EC2 instance types. Bandwidths of the virtual links are real numbers uniformly distributed between 1 and 50. VN's arrival times are generated randomly with arrival rate 10 VNs per 100 time units. The lifetimes of the VNRs are generated randomly between 300 and 700 time units with average 500 time units. 3000 VN topologies are generated and stored in brite format. For each algorithm, we run the simulation for 30000 time units with the previously generated VNRs. The generated SN topology, generated VNRs topologies, and outputs are available online at (Note 1).

For all algorithms, we set the maximum allowed hops (Max_hops) to 2, and the upper bound of remapping process (Max_backtrack) to *3n*, where *n* is the number of nodes in each VNR. We set the value of $CPU_{Max}$ in *Coarsening()* function (Algorithm 3) to the maximum available CPU in the sub-substrate network. Note that the BFSN-HEM algorithm does not guarantee that there are enough substrate nodes with available CPU equal to the $CPU_{Max}$ to embed coarsened virtual nodes if all virtual nodes are coarsened to nodes with CPU equal to $CPU_{Max}$, but our experiments have shown that it works very well. This is because, practically most of coarsened virtual nodes have CPU less than $CPU_{Max}$.

*5.2 Evaluation Results*

Three metrics have been used to evaluate the performance of the proposed algorithms: *the long-term average revenue*, which is defined by Equation (1), *the VNR acceptance ratio*, which is defined by Equation (2), and *the long-term R/Cost ratio*, which is defined by Equation (3). Figure 7 shows the simulation results using the VNR acceptance ratio to compare the different VNE algorithms. It can be seen that algorithms that embed VNs on best-fit sub-SNs increase the acceptance ratio compared with other algorithms. For example, at time unit 30000, in figure 7, the VNR acceptance ratio for the RW-BFS and RW-MaxMatch are 20 and 16 percent, while the VNR acceptance ratio for the BFSN and BFSN_HEM are 59 and 65 percent. Figure 7 also shows that the BFSN-HEM algorithm that coarsens VNs using *Heavy Edge Matching* technique has better VNR acceptance ratio than the BFSN algorithm. In other words, the proposed algorithms can embed more VNs on the same SN at the same time. Consequently, the proposed algorithms increase the long-term average revenue compared with other algorithms, as shown in figure 8. For example, at time unit 30000, the average revenue for the RW-BFS and RW-MaxMatch are 72 and 33, while the average revenue for the BFSN and BFSN_HEM are 290 and 336.

From figures 7 and 8, we can observe that the proposed algorithms enhanced the acceptance ratio around 40 percent while they tripled the long-term average revenue compared with other algorithms. This is because RW-BFS and RW-MaxMatch algorithms lead to SN's resource fragmentation. Thus, it may accept VNRs with small size and reject VNRs with large size. By accepting VNRs with large size, the average revenue of the BFSN and BFSN_HEM algorithms grows faster than the average revenue of the RW-BFS and RW-MaxMatch algorithms.

In addition to the higher revenue, the proposed algorithms are faster than other algorithms, because the proposed algorithms deal only with sub-SNs instead of the whole SN. Figure 9 depicts the *long-term Revenue/Cost ratio* for different VN embedding algorithms. As shown in figure 9, the long-term Revenue/Cost ratio of BFSN and BFSN-HEM are nearly the same, but the proposed algorithms perform slightly better than other algorithms. We believe that the reason behind this is the small value for the maximum allowed hops (Max_hops), which allows only embedding VNs if there are solutions with small cost.

## 6. Conclusion

In this paper, we proposed two virtual network embedding algorithms, which embed virtual networks on best-fit sub-substrate networks. Sub-substrate networks are specified by eliminating substrate links and substrate nodes that do not satisfy the connectivity and the CPU constraints. Virtualization technology is exploited to consolidated more than one virtual node from the same virtual network to one substrate node. Virtual networks are coarsened using *Heavy Edge Matching* (HEM) technique to minimize the cost of embedding virtual links. Our evaluation results show that the proposed algorithms improve the acceptance ratio and the revenue. Finally, we conclude that embedding virtual networks on best-fit sub-substrate networks allows substrate networks to accommodate more virtual networks. Additionally, coarsening virtual networks to coarsened virtual networks with minimal coarsened virtual links minimizes the embedding cost. For the future work, we plan to investigate other coarsening techniques to find the best coursing technique, which increases the acceptance ratio and the revenue while decreasing the embedding cost.

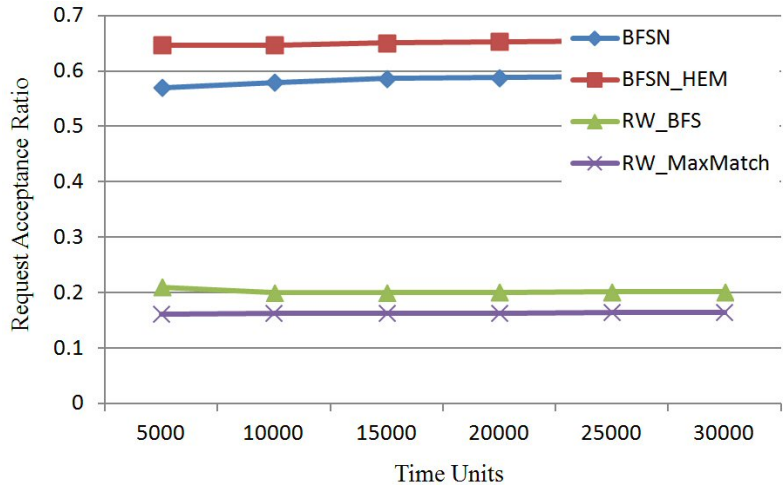

Figure 7. The VNR acceptance ratio comparison

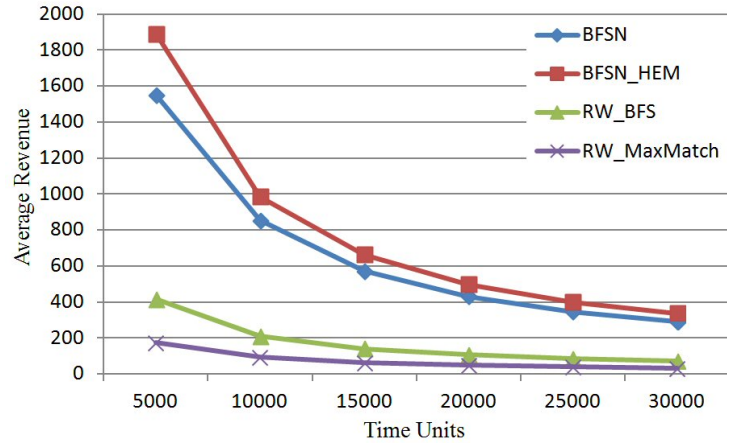

Figure 8. The long-term average revenue comparison

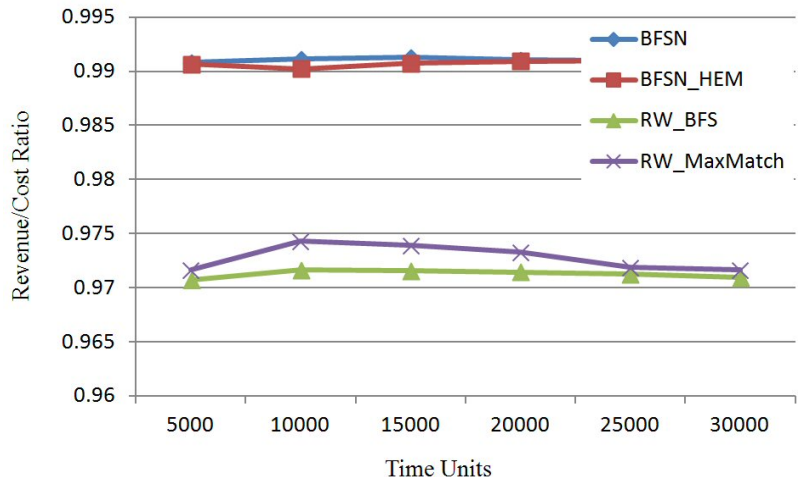

Figure 9. The long-term Revenue/Cost ratio comparison

**References**

Cheng, X., Su, S., Zhang, Z., Shuang, K., Yang, F., Luo, Y., & Wang, J. (2012). Virtual network embedding through topology awareness and optimization. *Computer Networks*, *56*(6), 1797-1813. http://dx.doi.org/10.1016/j.comnet.2012.01.022

Cheng, X., Su, S., Zhang, Z., Wang, H., Yang, F., Luo, Y., & Wang, J. (2011). Virtual network embedding through topology-aware node ranking. *SIGCOMM Comput. Commun. Rev.*, *41*(2), 38–47. http://doi.acm.org/10.1145/1971162.1971168

Chowdhury, M., Rahman, M., & Boutaba, R. (2012). Vineyard: Virtual network embedding algorithms with coordinated node and link mapping. *IEEE/ACM Transactions on Networking*, *20*(1), 206–219. http://dx.doi.org/10.1109/TNET.2011.2159308

Chowdhury, N., Rahman, M., & Boutaba, R. (2009). Virtual network embedding with coordinated node and link mapping. In *INFOCOM 2009, IEEE*, pages 783–791. http://dx.doi.org/10.1109/INFCOM.2009.5061987

Di, H., Li, L., Anand, V., Yu, H., & Sun, G. (2010). Cost efficient virtual infrastructure mapping using subgraph isomorphism. In *Communications and Photonics Conference and Exhibition (ACP), 2010 Asia*, pages 533–534. http://dx.doi.org/10.1109/ACP.2010.5682614

Fajjari, I., Aitsaadi, N., & Pujolle, G. (2013). Cloud networking: An overview of virtual network embedding strategies. In *Global Information Infrastructure Symposium, 2013*, pages 1–7. http://dx.doi.org/10.1109/GIIS.2013.6684379

Fajjari, I., Aitsaadi, N., Pujolle, G., & Zimmermann, H. (2011). VNR algorithm: A greedy approach for virtual networks reconfigurations. In *Global Telecommunications Conference (GLOBECOM 2011), 2011 IEEE*, pages 1–6. http://dx.doi.org/10.1109/GLOCOM.2011.6134006

Fischer, A., Botero, J., Till Beck, M., de Meer, H., & Hesselbach, X. (2013). Virtual network embedding: A survey. *Communications Surveys Tutorials, IEEE*, *15*(4), 1888–1906. http://dx.doi.org/10.1109/SURV.2013.013013.00155

Houidi, I., Louati, W., Ameur, W. B., & Zeghlache, D. (2011). Virtual network provisioning across multiple substrate networks. *Computer Networks*, *55*(4), 1011-1023. Special Issue on Architectures and Protocols for the Future Internet. http://dx.doi.org/10.1016/j.comnet.2010.12.011

Houidi, I., Louati, W., & Zeghlache, D. (2008a). A distributed and autonomic virtual network mapping framework. In *Fourth International Conference on Autonomic and Autonomous Systems, 2008. ICAS 2008*, pages 241–247. http://dx.doi.org/10.1109/ICAS.2008.40

Houidi, I., Louati, W., & Zeghlache, D. (2008b). A distributed virtual network mapping algorithm. In *IEEE International Conference on Communications, 2008. ICC '08*, pages 5634–5640. http://dx.doi.org/10.1109/ICC.2008.1056

Leivadeas, A., Papagianni, C., & Papavassiliou, S. (2013). Efficient resource mapping framework over networked clouds via iterated local search-based request partitioning. *IEEE Transactions on Parallel and Distributed Systems*, *24*(6), 1077–1086. http://dx.doi.org/10.1109/TPDS.2012.204

Lischka, J., & Karl, H. (2009). A virtual network mapping algorithm based on subgraph isomorphism detection. In *Proceedings of the 1st ACM Workshop on Virtualized Infrastructure Systems and Architectures*, VISA '09, pages 81–88, New York, NY, USA. ACM. http://dx.doi.org/10.1145/1592648.1592662

Lv, B., Wang, Z., Huang, T., Chen, J., & Liu, Y. (2010). Virtual resource organization and virtual network embedding across multiple domains. In *International Conference on Multimedia Information Networking and Security (MINES), 2010*, pages 725–728. http://dx.doi.org/10.1109/MINES.2010.154

Nogueira, J., Melo, M., Carapinha, J., & Sargento, S. (2011). Virtual network mapping into heterogeneous substrate networks. In *IEEE Symposium on Computers and Communications (ISCC), 2011*, 438–444. http://dx.doi.org/10.1109/ISCC.2011.5983876

Samuel, F., Chowdhury, M., & Boutaba, R. (2013). Polyvine: policy-based virtual network embedding across multiple domains. *Journal of Internet Services and Applications*, *4*(1). http://dx.doi.org/10.1186/1869-0238-4-6

Sun, G., Yu, H., Anand, V., & Li, L. (2013). A cost efficient framework and algorithm for embedding dynamic virtual network requests. *Future Generation Computer Systems*, *29*(5), 1265-1277. Special section: Hybrid Cloud Computing. http://dx.doi.org/10.1016/j.future.2012.08.002

Till Beck, M., Fischer, A., de Meer, H., Botero, J., & Hesselbach, X. (2013). A distributed, parallel, and generic virtual network embedding framework. In *IEEE International Conference on Communications (ICC), 2013*, 3471–3475. http://dx.doi.org/10.1109/ICC.2013.6655087

Xin, Y., Baldine, I., Mandal, A., Heermann, C., Chase, J., & Yumerefendi, A. (2011). Embedding virtual topologies in networked clouds. In *Proceedings of the 6th International Conference on Future Internet Technologies*, CFI '11, 26–29, New York, NY, USA. ACM. http://doi.acm.org/10.1145/2002396.2002403

Zhang, Z., Cheng, X., Su, S., Wang, Y., Shuang, K., & Luo, Y. (2013). A unified enhanced particle swarm optimization-based virtual network embedding algorithm. *Int. J. Communication Systems*, *26*(8), 1054–1073. http://dx.doi.org/10.1002/dac.1399

Zhu, Y., & Ammar, M. (2006). Algorithms for assigning substrate network resources to virtual network components. In *Proceedings of the 25th IEEE International Conference on Computer Communications*, INFOCOM 2006, 1–12. http://dx.doi.org/10.1109/INFOCOM.2006.322

**Note**

Note 1. Generated SN, generated VNRs, and outputs of the proposed algorithms are available at: https://drive.google.com/folderview?id=0BxEBmTQ0WG5RdGR1R2dOc3QyOE0&usp=sharing

**Copyrights**